\documentclass[prl,aps,amssymb,amsmath]{article}
\usepackage{amssymb,amsmath,tikz}
\usepackage{amsthm}
\usepackage{mathrsfs}
\usepackage{pifont}

\showoutput
\DeclareMathOperator*{\p}{p}

\newcommand{\ud}{\mathrm{d}}

\theoremstyle{plain}

\newtheorem*{twr*}{THEOREM}
\newtheorem*{lem*}{LEMMA}

\newtheorem*{rem*}{REMARK}

\newtheorem*{notn*}{NOTATION}
\newtheorem*{wiener-ito*}{WIENER-IT\^O-SEGAL DECOMPOSITION}

\begin{document}
\title{{\bf A commentary on single-photon wave function advocated by Bia{\l}ynicki-Birula}}
\author{Jaros{\l}aw Wawrzycki \footnote{Electronic address: jaroslaw.wawrzycki@wp.pl or jaroslaw.wawrzycki@ifj.edu.pl}
\\Institute of Nuclear Physics of PAS, ul. Radzikowskiego 152, 
\\31-342 Krak\'ow, Poland}
\maketitle

\vspace{1cm}

\begin{abstract}
We present in this paper how the single-photon wave function for transversal photons (with the direct sum of 
ordinary unitary representations of helicity 1 and -1
acting on it) is subsumed within the formalism of Gupta-Bleuler for the quantized free electromagnetic field. 
Rigorous Gupta-Bleuler quantization of the free electromagnetic field 
is based on our generalization (published formerly) of the Mackey theory of induced representations which includes representations preserving the indefinite Krein inner-product given by the Gupta-Bleuler operator. 
In particular it follows that the results of Bia{\l}ynicki-Birula on the single-photon wave function 
may be reconciled with the causal perturbative approach to QED.

\end{abstract}

This short account is a commentary on the single photon wave function as advocated by prof. Bia{\l}ynicki-Birula
(\cite{bialynicki-2} and references therein). His works on the subject enjoy a considerable attention and popularity.
This is because on the one hand the single photon wave function is a concept which is accompanied with
controversial opinions. Some authors, e.g. \cite{bohm}, even claim that position wave function 
for photon does not exist. But  on the other hand the subject being of fundamental importance, is still not 
systematically explored.   

We agree e.g. with \cite{Bialynicki} and \cite{bialynicki-2} and the authors cited there,
that the single photon wave function is already implicitly present in quantum field theory: 
generally a free quantum field is constructed by the application of
the symmetrized/antisymmetrized tensoring and direct sum operations (the so called second quantized functor) to a specific representation of the double covering of the Poincar\'e group acting in a space, which may be identified with the space of single particle wave functions,
and which depends on the specific quantum field. At the level of the free electromagnetic field one can 
start at the Hilbert space of transversal single photon states acted on by the direct sum of the unitary 
zero mass helicity 1 and $-1$ representations respectively (in the language of the classical by now Wigner-Mackey-Gelfand-Bargmann classification of irreducible unitary representations of the Poincar\'e group). In more physical terms the representation
has been described e.g in \cite{bialynicki-2} together with its relation to the Riemann-Silbertstein vector wave 
function. It is true that the (free) quantum electromagnetic field has its own peculiarities
making some differences in comparison to massive and non gauge fields which still serve as a source
of misunderstandings and still are not well understood. 

The first peculiarity of a zero mass quantum
(free) field, even non gauge field (as we assume for a while in order to simplify situation), is that now the representation
of the Poincar\'e group to which we apply Segal's functor of second quantization although being unitary in ordinary sense, is specified within the 
Wigner-Mackey classification scheme by the orbit in the momentum space which is the light cone (without the apex),
contrary to the massive case, where the orbit is the smooth sheet of the two-sheeted hyperboloid. The apex being a singular point of the cone (in the sense of the ordinary differential structure of the cone as embedded into the $\mathbb{R}^4$-manifold) causes serious difficulties of infra-red character. This is because the quantum field is 
in fact an operator-valued distribution (as motivated by the famous Bohr-Rosenfeld  analysis \cite{bohr-rosenfeld} of the measurement of the quantum electromagnetic field) which needs a test function space. It is customary to use the standard Schwartz space of rapidly decreasing
functions as the universal test space even for zero mass fields, and this is not the correct test space for zero mass field if it is supposed to be constructed with the help of annihilation-creation generalized operators (at fixed momenta) which have the rigorous meaning of white noise Hida operators. Note in particular that it is not the case for mass less field in the sense of Wightman which uses ordinary Schwartz test space, but his definition of field is useless in realistic causal perturbative QFT. On the other hand the white noise construction of free fields is crucial in the causal perturbative approach of Bogoliubov-Epstein-Glaser.
Recall that the construction mentioned to above of a free quantum field achieved by the second 
quantization functor $\Gamma$ applied to a representation specified by a 
fixed orbit allows to construct creation and annihilation families of \emph{ordinary operators} in the Fock space. In order to construct the field as operator valued distribution (or generalized operator in the white noise sense of Berein-Hida) we have to proceed much further then in the construction given by Streater and Wightman in their well known monograph \cite{wig}, Ch. 3. In the construction of Wightman we consider the restrictions of Fourier transforms (i.e. functions in the momentum space) of the test functions to the orbit in question. The construction works for the field construced throgh white noise Hida operators if the restriction is a continuous map from the test function space in $\mathbb{R}^4$ to the test function space in 
$\mathbb{R}^3$ which is really the case in the massive case as the orbit is a smooth manifold in that case. 
Unfortunately it seems that it has escaped due attention of physicists that 1) the white noise construction of free fields is crucial for the causal perturbative QFT and 2) that  the correct test function space
in the momentum representation for the zero mass field (if constructed with the help of Hida annihilation-creation operators) should be equal to the closed subspace $\mathcal{S}_0$ of the Schwartz space 
$\mathcal{S}$ of those functions
which vanish at zero together with all their derivatives and the test function space $\mathcal{S}_{00}$ in the position representation is given by the inverse Fourier image of the space $\mathcal{S}_0$. This in turn causes additional 
difficulties concerned with exploring and correct use of the principles of locality character, because in particular the
space $\mathcal{S}_{00}$ does not contain any function of compact support (except the trivial zero function) which immediately follows from the generalized Paley-Wiener theorem. But the splitting of causal distributions works still well 
because the pairing functions of free fields are homogeneous, and the test space $\mathcal{S}_{00}$ is flexible enough to provide the basis for the splitting of causal and homogeneous distributions into the retarded and advanced parts.
$\mathcal{S}_{00}$ is also flexible enough to distinguish conic-type subsets and in particular for the explorance 
of the casality relations needed for the perturbative construction of the scattering matrix in the causal approach
of Bogoliubov-Epstein-Glaser.

The situation for the electromagnetic field is still more delicate as the field is accompanied by the gauge freedom and the ordinary unitarity is untenable
and has to be replaced with a weaker condition of preservation of the indefinite Krein-inner product
-- which is the second main peculiarity of the electromagnetic field, shared with the other zero mass 
gauge fields of the standard model. This requires however the theory of non-unitary representations of the Poincar\'e group
which preserve indefinite inner product defined by the Gupta-Bleuler operator, which should allow us to work effectively with tensor products of such representations, Frobenius reciprocity theorem, 
imprimitivity system theorem, e.t.c.. Such a theory had not existed until 2015, compare \cite{wawrzycki-mackey}, where it appears for the first time. 

Therefore the construction of the field by the second quantization functor $\Gamma$
applied to a single particle representation should be extended on representations which are not unitary
but only Krein-isometric. 

Namely, although we may construct (remembering that we have to be careful with the choice of the test function space) the free quantum electric and magnetic fields by the mentioned 
application of the Segal second quantization functor to the direct sum of zero mas helicity 1 and $-1$ unitary representations acting on the Riemann-Silberstein vector function (as described in \cite{bialynicki-2}),
we encounter in this way a difficulty if we would like to restore the connection to the quantum vector potential
and its local transformation law within the the scheme. In principle we may reconstruct the quantum vector potential
in the momentum picture quite easily, but in connection to the non-local relationship of the potential to the electric and the magnetic fields in the position picture the local character of the vector potential is lost. 
We regard this a weakness when passing to interacting fields, specifically in passing to perturbative QED, and let us explain shortly why this is so. After half a century the causal method of St\"uckelberg and Bogoliubov turned up to be very valuable in avoiding the ultraviolet divergences in perturbative QFT, compare \cite{Epstein-Glaser}, 
\cite{epstein-glaser-al}. Their method have been extended on QED and the other gauge fields, compare \cite{BlaSen}, \cite{DKS1}, \cite{DKS2}, \cite{DKS3}, \cite{DKS4}. A crucial point of the method is the locality principle (local dependence of the interacting fields on the interaction Lagrangian, \cite{DutFred}), and the second circumstance is that we need to have the quantum vector potential -- recall that the minimal coupling is expressed immediately with the vector potential. Joining this prerequisites together we see that we need the vector potential with its local transformation law retained.
In particular we can achieve this within the Lorentz gauge with the four vector character of the transformation law of the quantum vector potential. However unitarity will have to be abandoned (recall the Gupta-Bleuler 
quantization \cite{Bleuler}).
Indeed, it is known that the four vector transformation law together with the zero mass character of the field cannot
be retained together with unitarity of the representation in the single particle space, compare e.g.
\cite{lop1}, \cite{lop2}. Because we prefer to stay within the micro-local perturbation scheme of QED and other gauge fields
of the standard model, avoiding ultraviolet divergences, we choose to abandon unitarity of the representation in the single photon states and replace Hilbert space and unitarity with Krein space and Krein-isometry property of the representation.
I.e. we  now have an ordinary Hilbert space together with two orthogonal projections $P_+$ and $P_-$
(of infinite dimensional ranges in our case) summing up to unity: $P_+ + P_- = I$, together with the 
fundamental symmetry $\mathfrak{J} = P_+ - P_-$
which in case of the Krein space of the free quantum electromagnetic field is called the Gupta-Bleuler operator $\eta$.
Exactly as the ordinary Hilbert space structure and unitary representation admits the operation 
of direct sum and tensoring also the Krein space structure and Krein isometric representation preserving
the Krein inner product $(\cdot, \mathfrak{J} \cdot)$ (where $(\cdot, \cdot)$ is the ordinary 
Hilbert space inner product) admits direct summation and tensoring. In order to work effectively with such Krein-isometric representations we need to built a theory which plays the role analogous to the Mackey theory of induced representations.
We have constructed such a theory in \cite{wawrzycki-mackey} and in particular we have proved the main theorems,
namely we proved the Kronecker product theorem, subgroup theorem and the imprimitivity system theorem to hold for Krein-isometric representations induced by Krein-unitary representations
 (additional analytic assumptions are sufficiently weak to be effective
for physical applications). As an application we construct the free quantum electromagnetic field using the 
symmetrized tensoring and direct summation to a specific single photon indecomposable (but reducible)
Krein-isometric representation of the double covering of the Poincar\'e group, which we we call 
\emph{{\L}opusza\'nski representation}. In short we apply the second quantization functor $\Gamma$
to the \emph{{\L}opusza\'nski representation} in order to construct the free electromagnetic field.
In particular the operator $\eta = \Gamma(\mathfrak{J})$, where $\mathfrak{J}$ is the fundamental symmetry of the 
single photon representation (i.e. \emph{{\L}opusza\'nski representation}), is indeed equal to the Gupta-Bleuler operator. 
 
It is the central assertion of this commentary that the construction of the transversal single photon space 
$\mathcal{H}_{\textrm{tr}}$ 
acted on by the  direct sum $[0,1] \oplus[0,-1]$ of unitary zero mass, helicty 1 and of helicity $-1$ representation as described in \cite{bialynicki-2} may be reconstructed as the closed subspace of physical states of the single photon 
Krein space with the representation $[0,1] \oplus[0,-1]$ induced by the action modulo unphysical states
of the single photon Krein-isometric representation. In this sense we extend the results on single photon wave
function, e.g. the uncertainty relations for the energy of single photon states obtained by prof. Bia{\l}ynicki-Birula
(compare \cite{bialynicki-2} and references therein) in showing their compatibility with the causal perturbative 
 QED, \cite{DutFred}, \cite{Bogoliubov_Shirkov}. For a more complete presentation of this point we refer 
to \cite{wawrzycki-photon}. Here we present only a brief account.

It should be stressed that the mathematically rigorous construction of the free gauge fields (e.g. quantum 
free electromagnetic potential) is not merely a matter of pedantry. In case of QED the ultraviolet
problem is fully solved by the extension of the Bogoliubov-Epstein-Glaser method \cite{Epstein-Glaser},
\cite{epstein-glaser-al} to QED, compare \cite{sharf}. The infrared divergences are controlled by the adiabatic 
switching of the interaction. However the infrared problem is only partially solved for QED in this way.
One aspect is that charged particles cannot be eigenstates of the mass operator. The other aspect are the divergences
which appear in the adiabatic limit. Here we comment shortly the second aspect (although it seems that these two aspects are interconnected). These divergences are logarithmic in QED and cancel out in the cross section, at least at lower order terms of the perturbative series \cite{sharf}. Blanchard and Seneor \cite{BlaSen} extended only partially on QED the result of Epstein-Glaser of the existence of the adiabatic limit for scalar massive field and proved the existence of the adiabatic limit for Wightman and Green functions for QED (for non-abelian gauge fields the situation is still less explored). In the Epstein-Glaser proof (for the scalar massive field) 
spectral condition is crucial, and essentially means that the orbit of the representation determining the single particle space is separated from zero and the only behaviour of the test function which plays a role goes through the restriction to the orbit of its Fourier transform (the test functions are just the Schwartz rapidly decreasing functions). Because the orbit  for the free electromagnetic field is not separated from zero being just the light cone, then the 
Epstein-Glaser proof doesn't work in QED. In the treatment of QED (and the other non-abelian gauge fields) 
we have not been so much pedantic in the construction of the free field,
because we have many relatively simple methods for making the correct guess as to the shape of 
distribution-functions giving the pairings of free fields plying the immediate role in computation of the cross section 
and even the Wightman functions for the free field. The successful solution of the ultraviolet problem and the cancellation of the infrared divergences 
in the cross section show that our guess of the pairings of the free fields was correct. 
But still something must go wrong when passing from cross section and c-number pairings to operator valued distributions
(generalized operators) themselves. Here enters our pedantry, because our rigorous construction of the zero mass gauge fields (e.g. the simplest free abelian gauge field -- i.e. the free electromagnetic field) revealed at least one point
which must have been missed at the heuristic level of the construction of the free field. Namely the test function space
has to be changed for the zero mass gauge fields, and in the momentum picture it is just $\mathcal{S}_0$. 
This in particular means that we have a God-given infrared cut-off assured by the very existence of the zero mass gauge field as a well defined operator valued distribution. In particular the method of Epstein-Glaser for the proof of the existence of the adiabatic limit should be revisited, because the fact that the light cone orbit is not separated from zero is compensated for by the infrared cut-off of the elements of $\mathcal{S}_0$. Another infrared problem which can be solved by the use of our rigorous construction is the strict proof of the Bogoliubov
quantization hypothesis for free fields,  as stated in \cite{Bogoliubov_Shirkov}. This problem lies among the problems which were unsolved and are concerned with the existence of integrals of local conserved currents corresponding to 
conserved symmetries, \cite{Requardt}, \cite{Maison-Reeh-1}, \cite{Maison-Reeh-2}. In case of zero mass gauge fields any endeavour of proving the existence of these integrals and their eventual equality to the generator of the corresponding one-parameter subgroup have permanently been accompanied by infrared divergences. Our rigorous method allows to solve these problems without encountering any divergences.     
      
After this general introduction let us concentrate on the main theme of our commentary and 
give some details of the single photon Krein-isometric representation and
the closed subspace of transversal photon states.
Let us start with a brief description of the Krein-isometric single photon {\L}opusza\'nski representation
in the momentum picture. We give at once the form of the representation which has the multiplier independent of the momentum, so that the Fourier transform of the momentum functions, i.e. position wave functions,
have local transformation formula. Namely the representation acts in a Krein space\footnote{We use the notation
of \cite{wawrzycki-photon}.} $(\mathcal{H}', \mathfrak{J}')$, i.e. an ordinary Hilbert space
$\mathcal{H}'$ endowed with the fundamental symmetry $\mathfrak{J}'^2 = I$, $\mathfrak{J}'^* = \mathfrak{J}'$,
and the Krein-isometric representation preserves the Krein-inner-product $(\cdot, \mathfrak{J}' \cdot)$, but
for detailed definition compare Sect. 2 of \cite{wawrzycki-mackey} as the peculiarities like unboundedness (with respect
to the ordinary Hilbert space product) cannot be excluded from the outset here in contrast to the ordinary unitary representations, and indeed our representation is unbounded. The Hilbert space $\mathcal{H}'$ consists 
of all measurable four component functions $\widetilde{\varphi}$ on the light cone $\mathcal{O}_{\bar{p}}$ in momentum space, which we may naturally regard 
as the functions of the spatial momentum components $\boldsymbol{\p} \in \mathbb{R}^3$ with $p^0(\boldsymbol{\p})
= r(\boldsymbol{\p})= \sqrt{\boldsymbol{\p} \cdot \boldsymbol{\p}}$, and which have finite Hilbert space norm
$\sqrt{(\cdot, \cdot)}$. The Hilbert space inner product $(\cdot, \cdot)$ in 
$\mathcal{H}'$ is equal 
\[
(\widetilde{\varphi}, \widetilde{\varphi}') 
= (\widetilde{\varphi}, B \widetilde{\varphi}')_{{}_{L^2(\mathbb{R}^3, \mathbb{C}^4)}} 
\]
where the self-adjoint positive operator  $B$, regarded as operator e.g. in $L^2(\mathbb{R}^3, \mathbb{C}^4)$, 
is equal to the operator of point wise multiplication by the  matrix operator 
\[
\frac{1}{2 r} B(p), \,\,\, p \in \mathcal{O}_{\bar{p}};
\]
which is strictly positive and self-adjoint in $\mathbb{C}^4$, with
\[
B(p) =  \\
\left( \begin{array}{cccc} 
\frac{r^{-2} + r^2}{2} & \frac{r^{-2} - r^2}{2r}p^1 & \frac{r^{-2} - r^2}{2r}p^2 & \frac{r^{-2} - r^2}{2r}p^3 \\
\frac{r^{-2} - r^2}{2r} p^1&\frac{r^{-2} + r^2 -2}{2r^2}p^1 p^1 +1 & \frac{r^{-2} + r^2 -2}{2r^2}p^1 p^2 & \frac{r^{-2} + r^2 -2}{2r^2} p^1 p^3  \\
\frac{r^{-2} - r^2}{2r}p^2 & \frac{r^{-2} + r^2 -2}{2r^2}p^2 p^1 &\frac{r^{-2} + r^2 -2}{2r^2}p^2 p^2 +1 & \frac{r^{-2} + r^2 -2}{2r^2} p^2 p^3  \\
\frac{r^{-2} - r^2}{2r}p^3 & \frac{r^{-2} + r^2 -2}{2r^2}p^3 p^1 & \frac{r^{-2} + r^2 -2}{2r^2}p^3 p^2 &\frac{r^{-2} + r^2 -2}{2r^2}p^3 p^3 +1  \end{array}\right), 
\]
again strictly positive self-adjoint on $\mathbb{C}^4$. For each $p \in \mathcal{O}_{\bar{p}}$ 
\begin{multline*}
{w_{{}_1}}^+(p) = \left( \begin{array}{c} 0 \\
                               \frac{p^2}{\sqrt{(p^1)^2 + (p^2)^2}}     \\
                             \frac{-p^1}{\sqrt{(p^1)^2 + (p^2)^2}}   \\
                               0       \end{array}\right), 
{w_{{}_1}}^- (p)= \left( \begin{array}{c} 0 \\
                               \frac{p^1 p^3}{\sqrt{(p^1)^2 + (p^2)^2}r}     \\
                             \frac{p^2 p^3}{\sqrt{(p^1)^2 + (p^2)^2}r}   \\
                              - \frac{\sqrt{(p^1)^2 + (p^2)^2}}{r}       \end{array}\right), \\ 
w_{{}_{r^{-2}}}(p) = \left( \begin{array}{c} \frac{1}{\sqrt{2}} \\
                              \frac{1}{\sqrt{2}}\frac{p^1}{r}     \\
                           \frac{1}{\sqrt{2}}\frac{p^2}{r}    \\
                             \frac{1}{\sqrt{2}}\frac{p^3}{r}       \end{array}\right),
w_{{}_{r^2}}(p) = \left( \begin{array}{c} \frac{1}{\sqrt{2}} \\
                              -\frac{1}{\sqrt{2}}\frac{p^1}{r}     \\
                           -\frac{1}{\sqrt{2}}\frac{p^2}{r}    \\
                             -\frac{1}{\sqrt{2}}\frac{p^3}{r}       \end{array}\right) 
\end{multline*} 
are the eigenvectors of the matrix $B(p)$ which are orthonormal in $\mathbb{C}^4$, where 
${w_{{}_1}}^+(p), {w_{{}_1}}^-(p)$ correspond to the eigenvalue equal $+1$, and 
$w_{{}_{r^{-2}}}(p), w_{{}_{r^2}}(p)$ correspond to the eigenvalues $r^{-2}, r^2$ respectively.

The fundamental symmetry $\mathfrak{J}'$ is equal to the operator of point wise multiplication 
by the matrix
\[
\mathfrak{J}'_{p} = \mathfrak{J}_{\bar{p}} B(p), \,\,\, p \in \mathcal{O}_{\bar{p}},
\]
with $\mathfrak{J}_{\bar{p}}$ equal to the following constant matrix
\[
\mathfrak{J}_{\bar{p}} = 
\left( \begin{array}{cccc} -1 & 0 & 0 & 0 \\
                                0 & 1 & 0 & 0     \\
                                0 & 0 & 1 & 0  \\
                                     0 & 0 & 0 & 1  \end{array}\right),
\]
being a fundamental symmetry in $\mathbb{C}^4$.  

If for each $\alpha \in SL(2, \mathbb{C})$ we denote by $\alpha \mapsto \Lambda(\alpha)$
the natural antihomomorphism of $SL(2, \mathbb{C})$ into the Lorentz group, and by 
$U(\alpha)$ the representors of $\alpha \in SL(2, \mathbb{C})$ and by $T(a)$,
$a \in \mathbb{R}^4$ the representors of translations, then we have 
\[
\begin{split}
U(\alpha) \widetilde{\varphi} (p) = \Lambda(\alpha^{-1}) \widetilde{\varphi} (\Lambda(\alpha)p), \\
T(a) \widetilde{\varphi}(p) 
= e^{i a \cdot p}\widetilde{\varphi}(p), \,\,\, \widetilde{\varphi} \in \mathcal{H}'.  
\end{split}
\]

The inverse Fourier transforms $\varphi$ 
\[
\varphi (x) = (2\pi)^{-3/2} \int \limits_{\mathscr{O}_{\bar{p}}} \widetilde{\varphi}(p) e^{-ip \cdot x} \, 
\ud \mu |_{{}_{\mathscr{O}_{\bar{p}}}} (p), \,\,\, \widetilde{\varphi} \in \mathcal{H}',
\]
compose the the single photon Krein space $(\mathcal{H}'', \mathfrak{J}'')$ in the position picture
with the representation giving the local four vector transformation law in the position picture.
In the last formula $\ud \mu |_{{}_{\mathscr{O}_{\bar{p}}}} (p)$ stands for the invariant measure
$2^{-1}r^{-1} \ud^3 \boldsymbol{\p}$ on the cone $\mathscr{O}_{\bar{p}}$.

Together with the {\L}opusza\'nski representation $(T,U)$ we consider the conjugate representation
$([T]^{*-1},[U]^{*-1}) = (\mathfrak{J}'T\mathfrak{J}', \mathfrak{J}' U \mathfrak{J}') = 
(T, \mathfrak{J}' U \mathfrak{J}')$, which likewise preserves the same Krein-inner-product
$(\cdot, \mathfrak{J}' \cdot)$.  
 
We apply to this conjugate representation the functor of second quantization obtaining the families $a(\widetilde{\varphi}),
a(\widetilde{\varphi})^+$ of creation and annihilation operators in the Fock space
$\Gamma(\mathcal{H}') \cong \Gamma(\mathcal{H}'')$, with the Gupta-Bleuler operator $\eta = \Gamma(\mathfrak{J}')$.
We claim that $\eta$ fulfils the correct commutation relations which are are to be expected 
for the Gupta-Bleuler operator.
   
We must be careful in preparing the fields as constructed with the hepl of white noise Hida operators. 
This can be achieved
by application of the Schwartz kernel theorem to the test function spaces $\mathcal{S}_{0}$
and $\mathcal{S}_{00}$ mentioned to above to the white noise generalized operator (operator valued distribution, quantum vector potential)
\[
A(\varphi) = A^\mu(\varphi_\mu)= a(\widetilde{\varphi}|_{{}_{\mathscr{O}_{\bar{p}}}}) 
+ \eta a(\widetilde{\varphi}|_{{}_{\mathscr{O}_{\bar{p}}}})^+ \eta,
\]
where $\varphi \in \mathcal{S}_{00}(\mathbb{R}^4)$, its Fourier transform $\widetilde{\varphi}$
belongs to $\mathcal{S}_{0}(\mathbb{R}^4)$ and where $\widetilde{\varphi} \mapsto 
\widetilde{\varphi}|_{{}_{\mathscr{O}_{\bar{p}}}}$ is the restriction to the cone, which turns out to be indeed 
a continuous map of nuclear spaces $\mathcal{S}_{0}(\mathbb{R}^4) \to \mathcal{S}_{0}(\mathbb{R}^3)$. 

It turns out that indeed the commutator
\[
[A(\varphi), A(\varphi')]
\] 
defines the kernel distribution equal to the Pauli-Jordan function multiplied by the minkowskian metric; 
and it follows that $A(\varphi)$ is the white-noise generalized operator, which is a local quantum field 
transforming locally as a four vector field.

It should be stressed that in general the elements $\widetilde{\varphi}$ of the single particle space of the
{\L}opusza\'nski representation (and its conjugation) in the momentum picture do not in general fulfil 
the condition $p^\mu \widetilde{\varphi}_\mu = 0$, so that in general their Fourier transforms
$\varphi$ do not preserve the Lorentz condition $\partial^\mu \varphi_\mu = 0$. This corresponds to the well known
fact that the Lorentz condition cannot be preserved as an operator equation. It can be preserved in the sense of 
the Krein-product average on a subspace of Lorentz states which arise from the closed subspace
$\mathcal{H}_{\textrm{tr}}$ of the so called transversal states together with all their images under the action
of the {\L}opusza\'nski representation and its conjugation. We are now going to define the closed subspace
$\mathcal{H}_{\textrm{tr}}$.

The closed subspace $\mathcal{H}_{\textrm{tr}} \subset \mathcal{H}'$ consists of all functions of the form
\[
\widetilde{\varphi} = {w_{{}_1}}^+ \, f_+ \, + \,
 {w_{{}_1}}^- \,f_-
\] 
with $f_+, f_-$ ranging over all pairs of measurable scalar functions on the light cone 
$\mathscr{O}_{\bar{p}}$ square integrable with respect to the invariant measure
$2^{-1}r^{-1} \ud^3 \boldsymbol{\p}$ on the cone. 
It follows that Hilbert space $\mathcal{H}'$ inner product 
\[
(\widetilde{\varphi}, \widetilde{\varphi}) =
\int \limits_{\mathscr{O}_{\bar{p}}} \, |f_+(p)|^2 \,\, 2^{-1}r^{-1} \ud^3 \boldsymbol{\p} 
+ \int \limits_{\mathscr{O}_{\bar{p}}} \, |f_-(p)|^2 \,\, 2^{-1}r^{-1} \ud^3 \boldsymbol{\p}
\] 
of any element $\widetilde{\varphi} \in \mathcal{H}_{\textrm{tr}}$
is equal to the Krein inner product $(\widetilde{\varphi}, \mathfrak{J}' \widetilde{\varphi})$, and thus the Krein
inner product is strictly positive on $\mathcal{H}_{\textrm{tr}}$. 

We claim that the action of the {\L}opusza\'nski representation and its conjugation generate modulo unphysical states of
Krein-norm zero and Krein orthogonal to $\mathcal{H}_{\textrm{tr}}$, exactly the same representation 
$(\mathbb{T}, \mathbb{U})$:
\[
\mathbb{U}(\alpha)\left( \begin{array}{c} f_{+} \\ 
                 f_{-} \end{array}\right)(p)
= \left( \begin{array}{cc} \cos \Theta(\alpha, p) & \sin \Theta(\alpha, p) \\ 
                          -\sin \Theta(\alpha, p) & \cos \Theta(\alpha, p)  \end{array}\right)
\left( \begin{array}{c} f_{+}(\Lambda(\alpha) p) \\ 
                 f_{-}(\Lambda(\alpha) p) \end{array}\right), 
\]
\[
\mathbb{T}(a)\left( \begin{array}{c} f_{+} \\ 
                 f_{-} \end{array}\right)(p)
= e^{ia\cdot p}
\left( \begin{array}{c} f_{+}(p) \\ 
                 f_{-}(p) \end{array}\right),
\]
on $\mathcal{H}_{\textrm{tr}}$, which is unitary for the strictly positive inner product on $\mathcal{H}_{\textrm{tr}}$
induced by the Krein-inner-product $(\cdot, \mathfrak{J}' \cdot)$, for the proof compare \cite{wawrzycki-photon}.
Therefore $(\mathbb{T}, \mathbb{U})$ is an ordinary unitary representation of the Poincar\'e group,
which may be shown to be unitary equivalent to the direct sum $[m=0, h=+1] \oplus [m=0, h=-1]$
of zero mass helicity $+1$ and of helicity $-1$ representations, \cite{wawrzycki-photon}.
For the concrete form of the phase $\Theta$ we refer to \cite{wawrzycki-photon}. The representation 
$(\mathbb{T}, \mathbb{U})$,
after a simple unitary transform on $\mathcal{H}_{\textrm{tr}}$, gives exactly the the single photon representation 
of \cite{bialynicki-2}, \S 4.3, formulas (4.22) and (4.23) with exactly the Hilbert space
of \S 5.1 of \cite{bialynicki-2},  which can be identified with our $\mathcal{H}_{\textrm{tr}}$,
compare \cite{wawrzycki-photon}.

\vspace*{0.2cm}

{\bf ACKNOWLEDGEMENTS}

The author is indebted for helpful discussions to prof. A. Staruszkiewicz.

%\section{Next section}
%The text...
%\subsection{Subsection}
%The text...

%uncomment the following lines to place a figure
%\begin{figure}[htb]
%\centerline{%
%\includegraphics[width=12.5cm]{Fig1}}
%\caption{Plot of ...}
%\label{Fig:F2H}
%\end{figure}

\end{document}